\documentclass[%
amsmath,amssymb,
aps, reprint
]{revtex4-2}
\usepackage{hyperref} 
\usepackage{graphicx}
\usepackage{dcolumn}
\usepackage{bm}
\usepackage{graphicx}
\usepackage{epstopdf}
\usepackage{mathrsfs}
\usepackage{amssymb}
\usepackage{slashed}
\usepackage{verbatim}
\usepackage{color}
\usepackage{multirow}
\usepackage{amsmath}
\usepackage{epsfig}
\usepackage{ulem}
\usepackage{graphics}
\usepackage{indentfirst}
\usepackage{slashed}
\newcommand{\bea}{\begin{eqnarray}}
\newcommand{\eea}{\end{eqnarray}}

\newcommand\nn{\nonumber}

\def\be{\begin{equation}}
\def\ee{\end{equation}}
\def\ba{\begin{eqnarray}}
\def\ea{\end{eqnarray}}

\def\<{\langle}
\def\>{\rangle}

\def\nn{\nonumber}

\def\cO {{\cal O}}

\newcommand{\Nc}{N_{\mathrm{crit}}}
\newcommand{\cM}{\mathcal{M}}
\newcommand{\cMad}{\mathcal{M}_{SU(N_f)\textrm{-ad}}}
\newcommand{\cMon}{\mathcal{M}_{SO(N_f^2-1)}}

\newcommand{\cTad}{\mathcal{T}_{\textrm{ad}}}

\newcommand{\nc}{\newcommand}
\newcommand{\ad}{\mathcal{O}_{\textrm{ad}}}
\newcommand{\Ad}{\textrm{Ad}}
\nc{\LL}{L}
\nc{\vv}{\tilde{v}}
\nc{\ccdot}{\!\cdot\!}
\nc{\gsm}{G_{SM}}
\nc{\vfive}{\mathbf{5}\oplus\mathbf{\overline{5}}}
\nc{\vten}{\mathbf{10}\oplus\mathbf{\overline{10}}}
\nc{\zhol}{Z^{\rm hol}}
\nc{\xfb}{\,{\rm fb}}

\begin{document}
 
\title{Bootstrapping conformal QED$_3$ and deconfined quantum critical point} 


\author{Zhijin Li}

\affiliation{Centro de F\'\i sica do Porto,
Departamento de F\'\i sica e Astronomia,
Faculdade de Ci\^encias da Universidade do Porto,
Rua do Campo Alegre 687,
4169--007 Porto, Portugal\\
Department of Physics, Yale University, New Haven, CT 06511, USA
}%


\begin{abstract}
     We bootstrap the deconfined quantum critical point (DQCP) and  3D Quantum Electrodynamics (QED$_3$) coupled to $N_f$ flavors of two-component Dirac fermions. We show the lattice and perturbative results on the $SO(5)$ symmetric DQCP are excluded by the bootstrap bounds combined with an irrelevant condition of the lowest singlet scalar. Remarkably, we discover a new family of kinks in the 3D $SO(N)$ vector bootstrap bounds with $N\geqslant6$.
     We demonstrate bound coincidences between $SU(N_f)$ adjoint and $SO(N_f^2-1)$ vector bootstrap which result from a novel algebraic relation between their crossing equations. By introducing gap assumptions breaking the $SO(N_f^2-1)$ symmetry, the $SU(N_f)$ adjoint bootstrap bounds with large $N_f$ converge to the $1/N_f$ perturbative results of QED$_3$. Our results provide strong evidence that the $SO(5)$ DQCP is not continuous and the critical flavor number of QED$_3$  is slightly above $2$: $N_f^*\in(2,4)$. Bootstrap results near $N_f^*$ are well consistent with the merger and annihilation mechanism for the loss of conformality in QED$_3$. 
\end{abstract}

\maketitle


\section{Introduction}  
The 3D Quantum Electrodynamics (QED$_3$) has been extensively studied in the past 30 years. In the low energy limit the theory becomes strongly coupled and provides an interesting laboratory to study confinement and chiral symmetry breaking. The infrared (IR) phase of QED$_3$ is determined by the number of fermions charged under the $U(1)$ gauge symmetry. The pure $U(1)$ gauge theory ($N_f=0$) confines \cite{Polyakov:1975rs, Polyakov:1976fu}. In the large $N_f$ limit QED$_3$ can be solved using $1/N_f$ expansion which gives an interacting stable IR fixed point \cite{Appelquist:1988sr}. The IR phase of QED$_3$ is quite subtle with small $N_f$:
the parity conserved mass of fermions could be generated dynamically and trigger spontaneously chiral symmetry
breaking \cite{Pisarski:1984dj, Appelquist:1986fd}\footnote{The chiral invariant but parity violating fermion mass could be generated dynamically as well, however, it costs more energy comparing with the parity invariant one so is less favored \cite{Appelquist:1986fd}.}.
There is a critical flavor number $N_f^*$ which separates the conformal phase from chiral symmetry breaking phase. 
It is of critical importance to determine $N_f^*$ for the applications of QED$_3$.
For instance, $N_f=4$ QED$_3$   has been applied to high-temperature cuprate superconductors  \cite{Rantner:2000wer,Rantner2002} and the order of phase transition is determined by $\Nc$.
Various approaches have been employed to estimate $N_f^*$ without a conclusive answer
\cite{Kubota:2001kk,  Kotikov:2016wrb,  Kaveh_2005, Giombi:2015haa, DiPietro:2015taa, Giombi:2016fct, Zerf:2018csr, Herbut:2016ide, Gusynin:2016som, benvenuti2019qeds, Braun:2014wja, Gukov:2016tnp, Hands:2002qt, Hands:2002dv, Hands:2004bh, Strouthos:2008kc, Karthik:2015sgq, Karthik:2016ppr}.

The $N_f=2$ QED$_3$ has been proposed to describe the deconfined quantum critical point (DQCP) \cite{Senthil:2003eed}.
A paradigmatic example of DQCP is the phase transition between N\'eel and Valence Bond Solid (VBS) phases of quantum antiferromagnets on the 2D square lattice. In the continuum limit, the N\'eel-VBS phase transition is described by the non-compact $CP^1$ (NCCP$^1$) model with $O(2)\times O(2)$  or $SO(3) \times U(1)$  symmetry, which are conjectured to be  dual to the $N_f=2$ QED$_3$ itself or coupled with a critical boson, i.e., the QED$_3$-GNY model.  The two theories are further conjectured to be self-dual with $O(4)/SO(5)$ symmetry enhancements and part of the 3D duality web \cite{Karch:2016sxi, Seiberg:2016gmd,  Wang:2017txt}, see \cite{Senthil:2018cru} for a review. This scenario has been carefully studied using lattice simulations \cite{Nahum:2015jya, Nahum:2015vka, Qin2017, Sreejith:2018ypg, Serna:2018tct, Xu:2018wyg}. There is promising evidence for symmetry enhancement while it is not clear whether the phase transition is continuous or weakly first order.
 
Modern conformal bootstrap \cite{Rattazzi:2008pe, Poland:2018epd} provides a powerful nonperturbative approach to study strongly coupled theories. Bootstrap studies on conformal QED$_3$ and DQCPs have been conducted in \cite{Chester:2016wrc, Chester:2017vdh, Nakayama:2016jhq, DSD, DP} which provide strict necessary conditions for the CFT data, though no clear evidence showing the bounds are saturated by conformal QED$_3$.
In this work, we  bootstrap  conformal QED$_3$ with an emphasis on the gauge invariant fermion bilinears. We start with the $SO(5)$ vector bootstrap. The bootstrap bounds provide strong constraints for the presumed $SO(5)$ symmetric DQCP. Remarkably, we discover a new family of kinks in the $SO(N)$ vector bootstrap bounds with $N\geqslant6$. We  study properties of these kinks and their connection with conformal QED$_3$.

\section{Bootstrap bounds on $SO(5)$ DQCP}
The $SO(5)$ symmetric DQCP has been suggested to be described by either  NCCP$^1$ model or $N_f=2$ QED$_3$-GNY model \cite{Wang:2017txt}. In NCCP$^1$ model, the $SO(5)$ vector multiplet contains the half-charged monopole  and N\'eel order parameter; while in $N_f=2$ QED$_3$-GNY model, it consists of a critical boson and the half-charged monopoles. Besides, the leading $SO(5)$ traceless symmetric scalar is constructed by the charge 1 monopoles and fermion bilinears in $N_f=2$ QED$_3$-GNY model or quadrilinear of the matter fields in NCCP$^1$ model. Above recipes of $SO(5)$ representations are helpful to test the emergent $SO(5)$ symmetry and dualities using lattice simulations \cite{Sandvik2007, Melko2008, Pujari2013, Nahum2015, Nahum:2015vka} or perturbative  approaches \cite{Dyer:2015zha, dupuis2021anomalous, benvenuti2019qeds, Boyack:2018zfx}.
The results provide promising evidence for the emergent $SO(5)$ symmetry and dualities. Nevertheless, the lattice  simulations in \cite{Nahum2015} observed drifting in the critical indices, indicating the IR phase is subtle. 

\begin{figure}
\includegraphics[width=0.9\linewidth]{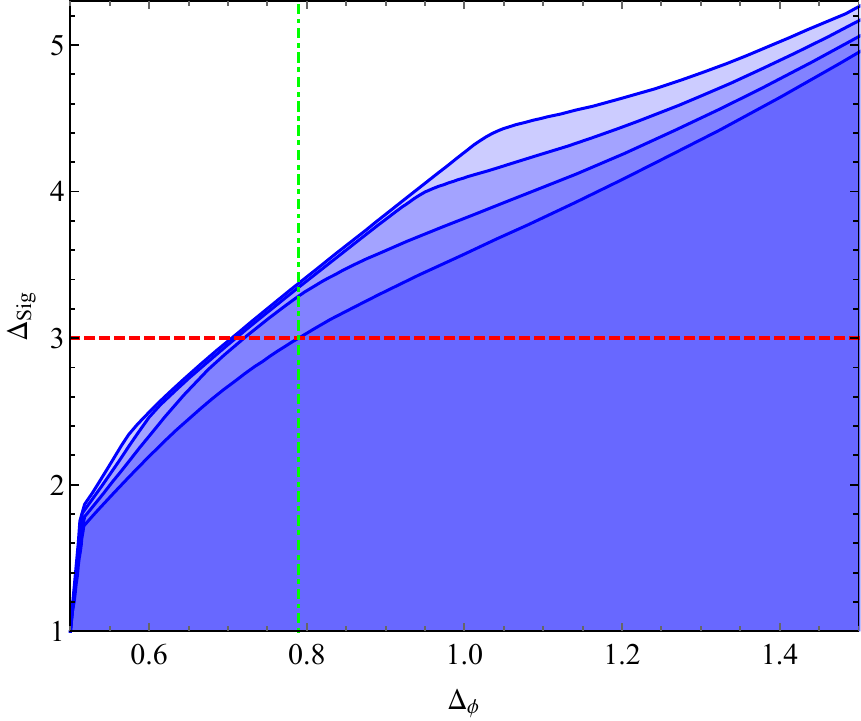}
 \begin{flushright}
\caption{Upper bounds ($\Lambda=31$) on the scaling dimensions of the $SO(N)$ singlet scalars. $N=5,6,7,8$ from bottom to top. The dot-dashed green line gives a left cut for the $SO(5)$ singlet bound with an irrelevant assumption $\Delta_{\textrm{Sig}}>3$.} \label{ONsig}
\end{flushright}
\end{figure}

\begin{figure}
\includegraphics[width=0.9\linewidth]{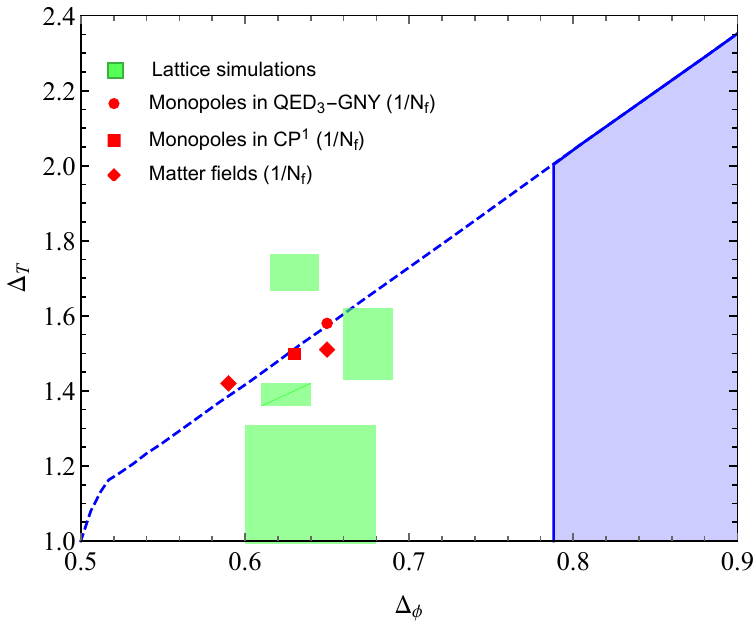}
 \begin{flushright}
\caption{Dashed blue line: upper bound on the  scaling dimension $\Delta_T$. Blue shadowed region: bootstrap allowed region of $(\Delta_\phi,\Delta_T)$ with an irrelevant assumption $\Delta_S>3$.} \label{O5T}
\end{flushright}
\end{figure}

In Figs. \ref{ONsig}-\ref{O5T} we show bootstrap bounds with $\Lambda=31$ \footnote{In bootstrap computations, $\Lambda$ is the order of the highest derivative in the linear functional, which determines the numerical precision  \cite{Simmons-Duffin:2015qma}. Unless specified explicitly, we will use $\Lambda=31$ throughout this paper.} on scaling dimensions of the lowest singlet ($\Delta_{\textrm{Sig}}$)  and traceless symmetric scalar ($\Delta_T$) in any unitary 3D $SO(5)$ symmetric CFTs. The bounds are smooth in most of the regions except sharp kinks in the left-bottom corner corresponding to the critical $O(5)$ vector model \cite{Kos:2013tga}. The lowest $SO(5)$ singlet scalar has to be irrelevant to realize the $SO(5)$ DQCP in lattice simulations without fine-tuning. This irrelevant condition leads to a lower cut on the $SO(5)$ vector scaling dimension $\Delta_\phi\geqslant0.79$. 
\begin{table}
\caption{CFT data of $SO(5)$ DQCP  ($=\textrm{est.}^{\textrm{err.}}$) estimated from lattice simulations or $1/N_f$ expansions. } \label{Table:Sig-T}
\begin{tabular}{cccccccc}
\hline\hline\\[-.5em]  Refs
			& \cite{Sandvik2007} & \cite{Melko2008}~ & ~\cite{Pujari2013}~ & ~\cite{Nahum2015} ~&  \cite{Dyer:2015zha} ~ &  \cite{dupuis2021anomalous} ~ &  \cite{Boyack:2018zfx}
 \\ [.5em]\hline \\[-.5em] $\Delta_\phi$
			&  $0.630^{15}$ & $0.675^{15}$~& $0.64^4$~& $0.625^{15}$ ~&  0.63 ~ &  0.65 ~ &  0.59/0.65 
 \\[.5em]\hline\\ [-.5em]  $\Delta_T$ 
			&  $1.716^{50}$ & $1.52^9$~& $1.11^{20}$~& $1.39^{3}$ ~&  1.50 ~ &  1.58 ~ &  1.42/1.51
 \\[.5em]\hline\hline 
		\end{tabular}
\end{table}
In Fig. \ref{O5T} we compared our bootstrap bounds with lattice simulations \cite{Sandvik2007, Melko2008, Pujari2013, Nahum2015, Nahum:2015vka} and perturbative  results \cite{Dyer:2015zha, dupuis2021anomalous, benvenuti2019qeds, Boyack:2018zfx} on the $SO(5)$ DQCP, which are summarized in Table. \ref{Table:Sig-T}.\footnote{ Note the critical indices from the lattice simulation \cite{Nahum2015} can drift to smaller values, e.g.  $\Delta_T\simeq 0.895^{85}$ with larger lattice sizes. The two sets of values in \cite{Boyack:2018zfx} are obtained from Pad\'e and Borel-Pad\'e approximations.} 
Most of the estimates on $\Delta_\phi$ locate in the range $(0.6,0.7)$, notably smaller than the lowest  bootstrap allowed 
value $\Delta_\phi\geqslant 0.79$. Therefore if the $SO(5)$ DQCP is described by a unitary CFT with an $SO(5)$ vector given in Table. \ref{Table:Sig-T},  there has to be a relevant singlet scalar. Such a relevant singlet scalar will necessarily affect the IR phase in the lattice simulations. The conclusion is that the phase transitions observed in above lattice simulations cannot be both $SO(5)$ symmetric and continuous.

However, bootstrap bounds in Figs.  \ref{ONsig} and \ref{O5T} do not exclude possible $SO(5)$ symmetric CFTs with $\Delta_\phi\in(0.6, 0.7)$ and a relevant singlet scalar. This scenario gets more intriguing considering that in Fig. \ref{O5T}, part of the lattice and $1/N_f$ results locate near the upper bound on $\Delta_T$ without the irrelevant assumption $\Delta_{\textrm{Sig}}>3$. It would be interesting to know whether the data near $SO(5)$ vector bootstrap bounds corresponds to a truly unitary  CFT.

A remarkable observation in Fig. \ref{ONsig} is that the $SO(N)$ singlet bounds show prominent kinks with irrelevant singlet scalars for $N\geqslant7$!\footnote{Actually there is another kink with a relevant singlet scalar in the $SO(8)$ vector bootstrap bound between the critical $O(8)$ vector model and the new family of kinks studied in this work. We will not study the property and the underlying physics of this kink. It will be very interesting  to study this kink carefully.} The kink becomes mild at $N=6$ and disappears at $N=5$, while the widely studied $SO(5)$ DQCP is just below the conformal window of this new family of kinks! Moreover, the kinks disappear accompanied by the lowest singlet scalar crossing the marginality condition $\Delta_\textrm{Sig}=3$. This is particularly interesting to study the loss of conformality \cite{Kubota:2001kk, Kaveh_2005, Gies:2005as, Kaplan:2009kr, Giombi:2015haa, Gorbenko:2018ncu} and we will discuss its possible interpretation later.

\begin{figure}
\includegraphics[width=0.9\linewidth]{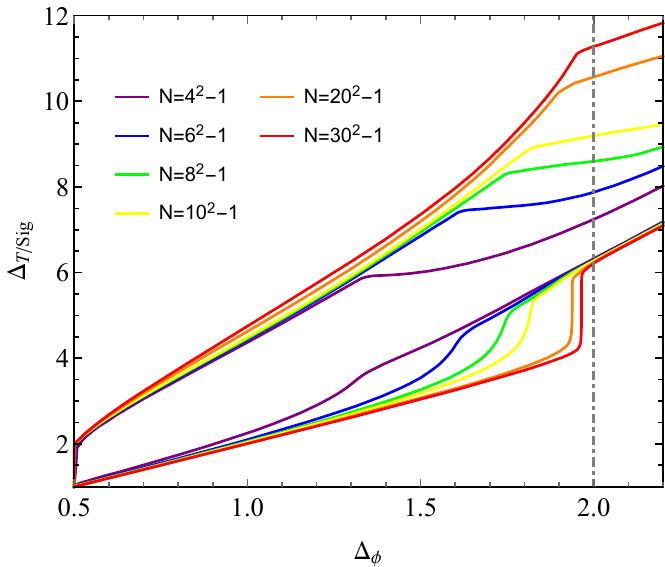}
\caption{Bounds on $\Delta_{\textrm{Sig}}$ (upper set) and $\Delta_T$ (lower set) from $SO(N)$ vector bootstrap.} \label{lgNs}
\end{figure}

\section{New family of kinks and QED$_3$}
\begin{figure}
\includegraphics[width=1\linewidth]{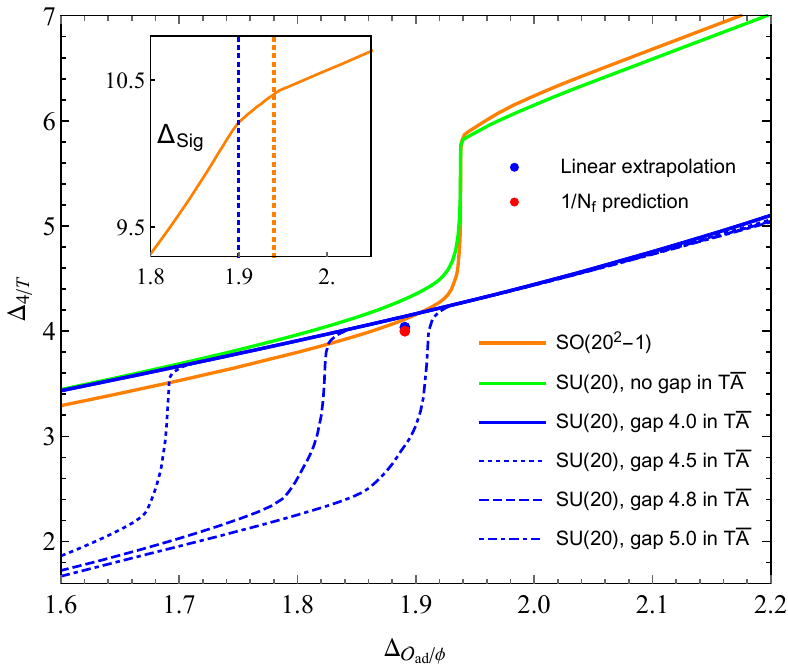}
 \begin{flushright}
\caption{Bounds on $\Delta_4$ and $\Delta_T$ from $SU(20)$ adjoint and $SO(399)$ vector bootstrap. In the left-top window, the orange line gives the singlet upper bound obtained from $SO(399)$ vector bootstrap with two kinks near $\Delta_\phi=1.90$ and  $1.94$.  } \label{SU20}
\end{flushright}
\end{figure}

To study underlying physics of the new family of kinks,
we start with their large $N$ behavior.
The $SO(N)$ vector bootstrap bounds with prominent kinks are shown in Fig. \ref{lgNs}. Actually the singlet bounds have interesting fine structures that there could be two nearby kinks, see Fig. \ref{SU20} and Appendix \ref{appendix:app2} for examples.\footnote{We remind the readers that our bootstrap bounds are not well-converged and the shapes of the bootstrap bounds could be affected by the numerical precision with different $\Lambda$s. It would be helpful to develop a more reliable bootstrap algorithm to determine the appearance and position of the kinks more efficiently and reliably. }
The kinks approach $\Delta_\phi=2$ in the large $N$ limit, which is exactly the scaling dimension of free fermion bilinears. A trouble to interpret the $SO(N)$ vectors as fermion bilinears is that the later carry zero or two symmetry indices and cannot be $SO(N)$ vectors. 
This puzzle relates to another surprising bootstrap phenomenon--coincidence of bounds from bootstrap with different symmetries. 

Bootstrap bound coincidence was firstly observed between singlet bounds from the $SO(2N)$ vector  and $SU(N)$ fundamental bootstrap  \cite{Poland:2011ey}. In our case, we find coincidence of singlet bounds from the $SU(N_f)$ adjoint and $SO(N_f^2-1)$ vector bootstrap. Such bound coincidence has also been noticed in \cite{Nakayama:2017vdd}.\footnote{Motivated by this work, a proof of the bootstrap bound coincidences between the $SU(N)$ fundamental and  $SO(N)$ vector bootstrap has been obtained in \cite{Li:2020bnb} which discovered a novel algebraic relation between the two crossing equations. The method can be directly generalized to more general examples \cite{li2020symmetries}. Moreover, in \cite{li2020symmetries} it shows the same algebraic relation can also be used to construct $SO(N)$ symmetric correlation functions from non-$SO(N)$ symmetric correlators.} Inspired by this fact, we conjecture that new kinks in the $SO(N)$ vector bootstrap bounds correspond to the gauge theories coupled with fermions, specifically the conformal QED$_3$, through $SO(N)$ symmetry enhancement in the bootstrap bounds.\footnote{In the non-gauged fermionic theories like the Gross-Neveu-Yukawa model, the fermion bilinears receive positive anomalous dimensions \cite{Gracey:2018fwq, Boyack:2018zfx}. In contrast the bootstrap results suggest that the $SO(N)$ vectors, if corresponding to fermion bilinears, should receive negative anomalous dimensions at finite $N$, consistent with the results in conformal gauge theories. Considering there are fine structures in the singlet bounds, it is possible that QED$_3$ is not the only theory that affects the bootstrap bounds, and other theories like QED$_3$-GNY model may also play a role here. While it needs more data to make this proposal clear. }
A clear understanding of the $SO(N)$ vector bootstrap kinks and their fine structures requires quantitatively precise CFT data which is beyond our current scope.\footnote{Bootstrap bounds near the kinks converge slowly. The large $\Lambda$ extrapolation method could be helpful to estimate the optimal bound of single point, while to determine the positions of the kinks,  it needs to know the shapes of the optimal bootstrap bounds which is much more difficult.} In this work, we study the conjecture from another direction: {\it if we break the $SO(N_f^2-1)$ symmetry in the $SU(N_f)$ adjoint bootstrap setup, will the bootstrap bounds converge to conformal QED$_3$?}
 
We bootstrap the four-point correlator of $SU(N_f)$ adjoint fermion bilinears in QED$_3$: ${\ad}_i^j\equiv\bar{\Psi}_i \Psi^j-\frac{1}{N_f}\delta_i^j\bar{\Psi}_k \Psi^k$. 
Crossing equations  for $SU(N_f)$ adjoint scalars $\cMad$  have been computed in \cite{Berkooz:2014yda, Iha:2016ppj}. Technical details and notations for the $SU(N_f)$ adjoint bootstrap are explained in Appendix \ref{appendix:app1}.
Remarkably, the $SU(N_f)$ adjoint crossing equations can be transformed to the $SO(N_f^2-1)$ vector crossing equations $\cMon$ through a linear transformation $\cTad$ \cite{Li:2020bnb, li2020symmetries}:
\be
\cTad\cdot \cMad \rightarrow \cMon, \label{SUtoSO}
\ee 
and the $SU(N_f)$ representations in $\cMad$ 
are mapped
to the $SO(N_f^2-1)$ representations
\bea 
SO(N_f^2-1) & \qquad\qquad &  SU(N_f) \nn \\
S &\longleftrightarrow & \mathbf{1}\;,   \label{branch1}\\
T &\longleftrightarrow & \Ad^+\oplus A\bar{A}\oplus  T\bar{T}\;,   \label{branch2}\\
A &\longleftrightarrow & \Ad^-\oplus T\bar{A}\;. \label{branch3}
\eea 
In conformal bootstrap, constraints on the CFT data are affected by two factors: crossing equations and gap assumptions. The algebraic relation (\ref{SUtoSO}) reveals that the $SU(N_f)$ adjoint and $SO(N_f^2-1)$ vector crossing equations actually have the same positive structure. If the gap assumptions do not break the $SO(N_f^2-1)\rightarrow SU(N_f)$ branching relations (\ref{branch1}-\ref{branch3}), the $SU(N_f)$ adjoint bootstrap degenerates to an $SO(N_f^2-1)$ vector bootstrap problem, see Appendix \ref{appendix:app1} for further explanation.

QED$_3$ spectrum breaks the $SO(N_f^2-1)$ symmetry from two aspects. Firstly, according to the branching rule (\ref{branch3}), the $SO(N_f^2-1)$ symmetry conserved current is decomposed into conserved currents in both $\Ad$ and $T\bar{A}$
sectors. However, in QED$_3$ the leading spin 1 operator in $T\bar{A}$ sector has scaling dimension $5\pm O(1/N_f)$, significantly above the unitary bound for large $N_f$. Secondly, in the three $SU(N_f)$ sectors branched from the $SO(N_f^2-1)$ $T$ sector (\ref{branch2}), the leading scalars in QED$_3$ are the four-fermion operators, whose scaling dimensions violate the $SO(N_f^2-1)$ symmetry at subleading order \cite{PhysRevB.78.054432,Chester:2016ref}:
{\small
\be
(\Delta_{\Ad}, \Delta_{A\bar{A}}, \Delta_{T\bar{T}} )\simeq
(4-\frac{185}{3\pi^2N_f}, 
 4-\frac{64}{\pi^2N_f}, 
 4+\frac{64}{3\pi^2N_f}). \label{4fm}
\ee }

To obtain bootstrap bounds for  conformal QED$_3$, it is necessary to introduce gap assumptions breaking the $SO(N_f^2-1)$ symmetry. While it remains a highly non-trivial question whether these gap assumptions are sufficient to generate bootstrap bounds nearly saturated by conformal QED$_3$.
To answer this equation, we bootstrap the $N_f=20$ conformal QED$_3$, for which the $1/N_f$ expansions at subleading order are expected to be close to the physical spectrum. We introduce gap assumptions in $\Ad_{\ell=0}, A\bar{A}_{\ell=0}, T\bar{T}_{\ell=0}$ sectors which break the $SO(N_f^2-1)$ symmetry as the $1/N_f$ corrections in (\ref{4fm}):
\be
\Delta\geqslant (\Delta_4-\frac{185}{3\pi^2N_f},\Delta_4-\frac{64}{\pi^2N_f}, \Delta_4+\frac{64}{3\pi^2N_f}  )\label{gaps1}
\ee
and bootstrap upper bound on $\Delta_4$.
 In the physical spectrum of  QED$_3$ with large $N_f$, we have $\Delta_4\simeq4$.

\begin{table}
\caption{Linear extrapolations of the upper bounds on $\Delta_4$ ($\simeq4$ in QED$_3$) with $\Delta_{\ad}$ fixed at the $1/N_f$ results ($\Delta_1^*=4$).  \label{slrule}
}
 
\begin{tabular}{cccccccc}
\hline\hline\\[-.5em] ~ $N_f$ ~
			& ~10~ & ~20~ & ~30~ & ~50 ~&  100 ~ &  150 ~ &  200 ~
 \\[.5em]\hline\\ [-.5em] ~ $\Delta_{4}$ ~
			& ~ 4.083~ & 4.038~& 4.024~& 4.017 ~&  4.005 ~ &  4.004 ~ &  4.001~
 \\[.5em]\hline\hline 
		\end{tabular}
\end{table}

The $SU(20)$ adjoint bootstrap bounds on $\Delta_4$ are shown in Fig. \ref{SU20}.  
There are three interesting properties: $\mathbf{a}$, the first kink in the singlet bound corresponds to the bottom of the jump in the bound on $\Delta_T$ (orange line), while the top of the jump in the bound on $\Delta_T$ corresponds to the second mild kink in the singlet bound; $\mathbf{b}$, bounds on $\Delta_4$ with  gaps (\ref{gaps1}) (green line) are shifted in opposite directions from $\Delta_T$ bound before and after the jump, while $\Delta_4\simeq\Delta_T$ at the top of the jump; $\mathbf{c}$, the gap $\Delta_1^*$ is particularly interesting. Gaps $\Delta_1^*=4.5,4.8,5.0$ generate sharp jumps in the bounds on $\Delta_4$, while bootstrap bounds right to the jumps are not sensitive to the specific values of $\Delta_1^*$! With gap $\Delta_1^*=5.0$ the large $N_f$ predictions on QED$_3$ (red dot) is excluded while the gap $\Delta_1^*=4.8$ generates a jump near the physical value $\Delta_4=4$.
Coefficient of the $1/N_f$ term in $\Delta_1^*=5\pm O(1/N_f)$ is not known yet but our results suggest the subleading order correction should be negative. See Appendix \ref{gapinTAb} for more discussions.
Using linear extrapolation of the upper bounds on $\Delta_4$ with gaps (\ref{gaps1}) and $\Delta_1^*=4$, it gives an optimal upper bound $\Delta_4\simeq 4.038$ near $\Delta_{\ad}\simeq1.891$, remarkably close to the physical value $\Delta_4=4$! The small discrepancy could be explained by higher order corrections to the CFT data in (\ref{4fm}).\footnote{The subleading order corrections in (\ref{4fm}) are at the order $O(10^{-1})$. One may expect the next-to-subleading order corrections at the order $O(10^{-2})$, comparable to the discrepancy. Note the results may also be affected by the systematical errors from linear extrapolation.}  Bootstrap bounds on central charges in Appendix \ref{appendix:app3} are  especially close to the large $N_f$ predictions of QED$_3$ near $\Delta_4=4$, which suggest the bootstrap solutions related to 3D Abelian gauge theories instead of  Yang-Mills gauge theories. In Table \ref{slrule} we show more comparisons between perturbative results and linear extrapolations of bounds on $\Delta_4$.\footnote{Bootstrap results with $\Lambda=19,21,...,35$ used in the linear extrapolations are provided in an attached {\it Mathematica} file. We used binary search to compute upper bounds on $\Delta_4$ with numerical precision $10^{-5}$.}
Agreements between the two methods get more impressive with increasing $N_f$. 
The results provide strong evidence for the question we want to address: {\it the $SU(N_f)$ adjoint bootstrap bounds, after resolving the $SO(N_f^2-1)$ symmetry enhancement, are close to be saturated by conformal QED$_3$!}

Let us go back to the observation in Fig. \ref{ONsig} that the kinks disappear with the lowest singlet scalar crossing marginality condition. According to the proposed relation between the new kinks and conformal QED$_3$, disappearance of the kinks relates to loss of conformality in QED$_3$, indicating a critical flavor number slightly above 2: $N_f^*\in(2,4)$. 
In consequence, the DQCPs related to $N_f=2$ QED$_3$ or QED$_3$-GNY
model may violate unitarity by a small complex factor, which could explain the quasi-conformal behavior observed in lattice simulations \cite{Gorbenko:2018ncu, Gorbenko:2018dtm}.
The marginally irrelevant scalar near $N_f^*$ can be nicely interpreted by the merger and annihilation mechanism for the loss of conformality in QED$_3$  \cite{Kaplan:2009kr, Gorbenko:2018ncu}. In this scenario, the lowest singlet scalar becomes relevant below $N_f^*$ which generates a RG flow dissolving the IR fixed points. A critical prediction of this mechanism is the singlet scalar approaching marginality condition $\Delta_{\textrm{Sig}}=3$ from above near $N_f^*$, which is surprisingly consistent with the behavior of kinks in Fig. \ref{ONsig}.\footnote{To confirm the merger and annihilation mechanism further, it needs to measure the index $a$ in the relation $\Delta_{\textrm{Sig}}-3\sim (N_f-N_f^*)^a$. Since the singlet bounds converge slowly and there is an $SO(N_f^2-1)$ symmetry enhancement in the bootstrap bounds, it is hard to estimate this index reliably based on our current results.}

\section{Discussions}
In this work we have bootstrapped the $SO(5)$ symmetric DQCP and our results suggest the phase transitions observed in previous lattice simulations cannot be both $SO(5)$ symmetric and continuous.
Moreover, we discovered a new family of  kinks in the $SO(N)$ vector bootstrap bounds
with $N\geqslant6$, while the  $SO(5)$ DQCP is just slightly below the window. These kinks show interesting fine structures which require more  bootstrap data for a clear understanding.
We observed bound coincidences between the $SU(N_f)$ adjoint and $SO(N_f^2-1)$ vector bootstrap and explained that this is caused by an algebraic relation between the two crossing equations. We have shown that for general large $N_f$s, with gaps breaking $SO(N_f^2-1)$ symmetry the $SU(N_f)$ adjoint bootstrap bounds converge to conformal QED$_3$. Our results support the merger and annihilation mechanism for the loss of conformality in QED$_3$, and indicate a critical flavor number of QED$_3$: $N_f^*\in(2,4)$.


Our results are illuminating for the widely interested project on solving conformal QED$_3$ with bootstrap. On the one hand, our results indicate the CFT landscape is not tameless--after introducing suitable $SO(N)$ symmetry breaking gaps the bootstrap bounds indeed get close to the conformal QED$_3$. Bootstrap can generate highly nontrivial constraints on the theory, especially in the strongly coupled region with small $N_f$,  e.g. the $SO(5)$ DQCP. On the other hand, our results also clarified that to numerically solve conformal QED$_3$ with bootstrap, a crucial challenge is to resolve the  $SO(N)$ symmetry enhancement in the crossing equations and reproduce proper spectrum of QED$_3$, for which certain substantially new ingredients  are needed in conformal bootstrap.

We have observed similar family of kinks in the $SO(N)$ vector bootstrap bounds in general spacetime dimension $D$ with a critical flavor number increasing with $D$. The kinks in 4D are especially interesting-- they are also deformed from 4D free fermion bilinears and it would be extremely interesting to know what we can learn from them about the profound 4D gauge theories \cite{Caswell:1974gg, Banks:1981nn}. 
 
\section{acknowledgments}
The author thanks Miguel Costa, Zohar Komargodski, David Poland,  Junchen Rong, Slava Rychkov, Christopher Pope, Ning Su  and Cenke Xu for valuable discussions.  The author is especially grateful to Christopher Pope for his consistent support throughout the course of this work. The author is greatly benefited from collaborations with Ning Su on numerical conformal bootstrap.
The author would like to thank the organizers of the ``Project Meeting: Analytical Approaches workshop at
Azores and the  Bootstrap 2018 workshop at Caltech, for their hospitality and
support. This research received funding from the grant CERN/FIS-PAR/0019/2017.
This work was also supported by the
Simons Foundation grant 488637 and 488651 (Simons collaboration on the Nonperturbative bootstrap) and the DOE grant no. DE-SC0020318.
Centro de F\'isica do Porto is partially funded by the Foundation for Science and Technology
of Portugal (FCT).
The computations in this work were run on the Mac Lab cluster supported by the Department of Physics and Astronomy,  Texas A\&M University, and the Yale Grace computing cluster, supported by the facilities
and staff of the Yale University Faculty of Sciences High Performance Computing Center.


\bibliography{DQCP-QED3}

\clearpage

\onecolumngrid
 \begin{center}
{\bf {\large Supplementary material}}
\end{center}

\appendix

\section{$SU(N_f)$ adjoint bootstrap and its relation with the $SO(N_f^2-1)$ vector bootstrap}\label{appendix:app1} 

In this section we show a novel algebraic relation between crossing equations of the $SU(N_f)$ adjoint  and $SO(N_f^2-1)$ vector scalars. We will  follow the methods developed in \cite{Li:2020bnb, li2020symmetries} which were motivated by the new family of kinks in the $SO(N)$ vector bootstrap bounds discovered in this work. Considering the special role of this relation in determining the $SU(N_f)$ adjoint bootstrap bounds, we present it in the new version of this work to explain the $SO(N_f^2-1)$ symmetry enhancement in the $SU(N_f)$ adjoint bootstrap bounds and study their relation with conformal QED$_3$ proposed in the first version of this paper.

Let us consider the four-point correlator of an $SU(N_f)$ ($N_f\geqslant 4$) adjoint scalar $\ad$ 
\be
\langle\ad(x_1)\ad(x_2)\ad(x_3)\ad(x_4)\rangle. \label{4pt}
\ee
There are six representations appearing in its s or t-channel conformal partial wave expansions, corresponding to the OPE
\be
\ad\times \ad \rightarrow 
\mathbf{1}^+\oplus \Ad^+\oplus\Ad^-\oplus A\bar{A}^+\oplus (T\bar{A}+A\bar{T})^-\oplus T\bar{T}^+,  \label{OPE}
\ee
where the $\mathbf{1}$ and $\Ad$ denote the singlet and adjoint representations of $SU(N_f)$. $A/T$ ($\bar{A}/\bar{T}$) denote representations with anti-symmetric/symmetric fundamental (anti-fundamental) indices of $SU(N_f)$. Moreover, operator $\ad$
is real, so are the operators in its OPE, therefore only the real combination of representation $T\bar{A}$ and its complex conjugation $A\bar{T}$ can appear in (\ref{OPE}). We will denote this sector by $T\bar{A}$ for simplicity. The superscripts in $R^\pm$ denote even/odd spin selection rules in the representation $R$. 

Crossing equations of the four-point correlator (\ref{4pt})  have been obtained in previous bootstrap studies \cite{Berkooz:2014yda, Iha:2016ppj}, which can be written in a compact form
\be 
\label{eq:abcreq}
\sum_{\cO\in\ell^+} \lambda_\cO^2\Vec{V}^+_{\mathbf{1}}+\sum_{\cO\in\ell^+} \lambda_\cO^2\Vec{V}^+_{\Ad}+\sum_{\cO\in\ell^-} \lambda_\cO^2\Vec{V}^-_{\Ad}\\+\sum_{\cO\in\ell^-} \lambda_\cO^2\Vec{V}^-_{T\bar{A}}+\sum_{\cO\in\ell^+} \lambda_\cO^2\Vec{V}^+_{A\bar{A}}+\sum_{\cO\in\ell^+} \lambda_\cO^2\Vec{V}^+_{T\bar{T}}=0\;.
\ee 
Here the vector $\Vec{V}^\pm_{R}$ corresponds to the $SU(N_f)$ representation $R$ with even/odd spins. Explicit form of each vector can be summarized in the matrix
\bea
\cMad=\left(
\begin{array}{cccccc}
 0 & 0 & 0 & -F & F & F \\
 0 & \frac{2 F}{N_f} & 0 & 0 & -\frac{F}{N_f-2} & \frac{F}{N_f+2} \\
 0 & -F & -F & \frac{F}{N_f} & \frac{F}{N_f-2} & \frac{F}{N_f+2} \\
 F & -\frac{16 F}{N_f} & 0 & 0 & \frac{ 2 N_f^2F}{(N_f-1) (N_f-2)} & \frac{2 N_f^2 F}{(N_f+1) (N_f+2)} \\
 H & -\frac{4 H}{N_f} & 0 & -H & -\frac{N_f (N_f-3)H}{(N_f-1) (N_f-2)} & -\frac{N_f (N_f+3)H}{(N_f+1) (N_f+2)} \\
 0 & H & -H & \frac{H}{N_f} & \frac{(N_f-3)H}{N_f-2} & -\frac{ (N_f+3)H}{N_f+2} \\
\end{array}
\right),
\eea
in which 
\bea 
F &= v^{\Delta_{\ad}}g_{\Delta,\ell}(u,v)-u^{\Delta_{\ad}}g_{\Delta,\ell}(v,u)\;, \label{Fblock}\\
H &= v^{\Delta_{\ad}}g_{\Delta,\ell}(u,v)+u^{\Delta_{\ad}}g_{\Delta,\ell}(v,u)\;.
\eea
are the (anti-)symmetrized conformal block functions.
Columns in $\cMad$ give the vectors $\Vec{V}^\pm_{R}$ in the order
\be 
\cMad=\left( \Vec{V}^+_{\mathbf{1}}, \Vec{V}^+_{\Ad}, \Vec{V}^-_{\Ad},
\Vec{V}^-_{T\bar{A}},  \Vec{V}^+_{A\bar{A}}, \Vec{V}^+_{T\bar{T}}\right)\,.
\ee 
Similarly the $SO(N)$ vector crossing equations can be written in a matrix form
\be
 \cM_{SO(N)}= \left(\Vec{V}^+_S, \Vec{V}^+_T, \Vec{V}^-_A   \right)_{SO(15)}=
\left(
\begin{array}{ccc}
 0 & F & -F \\
 F & (1-\frac{2}{N})F & F \\
 H & -(1+\frac{2}{N}) H & -H \\
\end{array}
\right)\,. \label{SOceq}
\ee
It has been shown in \cite{Li:2020bnb, li2020symmetries} that the two crossing equations can be connected through a linear transformation 
\be
\cTad=\left(
\begin{array}{cccccc}
 1 & \frac{2 \left(N_f^4-2 N_f^2+2\right)}{N_f^4-N_f^2-2} & \frac{2 N_f}{N_f^2-2} & 0 & 0 & 0 \\
 -1 & \frac{-8 N_f^4+16 N_f^2+4}{-N_f^4+N_f^2+2} & -\frac{2 N_f}{N_f^2-2} & 1 & 0 & 0 \\
 0 & 0 & 0 & 0 & 1 & \frac{2 N_f}{N_f^2-2} \\
\end{array}
\right),
\ee
which maps the  crossing equations $\cMad$ to the crossing equations $\cMon$:
\bea 
\cTad \cdot \cMad &=& 
\left(
\begin{array}{cccccc}
 0 & x_1F & - x_2F & -x_3F & x_4F & x_5F \\
 F &  x_1F \left(1-\frac{2}{N_f^2-1}F\right) & x_2F & x_3F & x_4F \left(1-\frac{2}{N_f^2-1}\right)F & x_5F \left(1-\frac{2}{N_f^2-1}\right) \\
 H &  -x_1H \left(1+\frac{2}{N_f^2-1}\right)H & -x_2H & -x_3H & -x_4 \left(1+\frac{2}{N_f^2-1}\right)H & -x_5 \left(1+\frac{2}{N_f^2-1}\right)H \\
\end{array}
\right)  \nn\\
&=&\left( \Vec{V}^+_{S},\; x_1\Vec{V}^+_{T},\; x_2\Vec{V}^-_{A},\;
x_3\Vec{V}^-_{A},\; x_4\Vec{V}^+_{T},\; x_5\Vec{V}^+_{T}\right)\,, \label{TMSU}
\eea
with positive coefficients $\vec{x}$
\be
\vec{x}=\left\{ \frac{2 \left(N_f^4-5 N_f^2+4\right)}{N_f \left(N_f^4-N_f^2-2\right)},  ~\frac{2 N_f}{N_f^2-2},  ~1-\frac{2}{N_f^2-2},~  \frac{\left(N_f-3\right) N_f^2 \left(N_f+1\right)}{\left(N_f^2-2\right) \left(N_f^2+1\right)}, ~ \frac{\left(N_f-1\right) N_f^2 \left(N_f+3\right)}{\left(N_f^2-2\right) \left(N_f^2+1\right)}\right\}.
\ee
The right part of (\ref{TMSU}) 
is just the $SO(N_f^2-1)$ vector crossing equations $\cMon$ associated with the $SO(N_f^2-1)\rightarrow SU(N_f)$ branching rules given by (\ref{branch1}-\ref{branch3}).
The algebraic relation (\ref{TMSU}) connects the $SU(N_f)$ adjoint bootstrap problem with the $SO(N_f^2-1)$ vector bootstrap problem in the following way.

Assume we have obtained linear functionals $\Vec{\alpha}\equiv (\alpha_1,\, \alpha_2, \, \alpha_3)$ for the $SO(N_f^2-1)$ vector bootstrap, i.e.,
\be
\label{actMon}
\Vec{\alpha}\cdot \cMon=\Vec{\alpha}\cdot\left(\Vec{V}^+_S, \Vec{V}^+_T, \Vec{V}^-_A   \right)= (\alpha_{S}^+, \, \alpha_{T}^+,\, \alpha_{A}^-)\succeq 0_{1\times 3}\;,\hspace{1cm} \forall \Delta\geqslant \Delta_{S/T/A, \ell}^*\;, 
\ee 
then the linear functionals $\Vec{\alpha}$ could be used to construct linear functionals for the $SU(N_f)$
adjoint bootstrap. Specifically, the action of the linear functionals 
$
\Vec{\beta} =\Vec{\alpha} \cdot \left(\cTad\right)
$ on the  $SU(N_f)$ adjoint bootstrap equations is
\bea 
\Vec{\beta}\cdot \cMad&=&\left(\Vec{\alpha}\cdot \cTad\right)\cdot\cMad =\Vec{\alpha}\cdot \left( \Vec{V}^+_{S}, \, x_1\Vec{V}^+_{T},\, x_2\Vec{V}^-_{A},\,
x_3\Vec{V}^-_{A}, \, x_4\Vec{V}^+_{T},\, x_5\Vec{V}^+_{T}\right)  \\
&=&\left( \alpha_{S}^+, \, x_1\alpha_{T}^+,\, x_2\alpha_{A}^-,\,x_3\alpha_{A}^-, \, x_4\alpha_{T}^+,\, x_5\alpha_{T}^+  \right), \hspace{1cm} \forall \Delta\geqslant \Delta_{R_i, \ell}^*\,. \label{actMad}
\eea 
As long as the gap assumptions $\Delta_{R_i, \ell}^*$ in (\ref{actMad}) is consistent with the gap assumptions in (\ref{actMon}), the linear functional actions in (\ref{actMad}) also satisfy the positive conditions. This leads to a conclusion that any linear functionals that can be used to exclude the CFT data in $SO(N_f^2-1)$
vector bootstrap can also be used to exclude the CFT data in $SU(N_f)$ adjoint bootstrap. Also any $SO(N_f^2-1)$ symmetric solutions to the crossing equations can be decomposed into the solutions of the $SU(N_f)$ adjoint crossing equations. Therefore the bootstrap allowed regions of the two different bootstrap setup are actually exactly the same! If one adopts different gap assumptions which are not consistent with the $SO(N_f^2-1)\rightarrow SU(N_f)$ branching rules (\ref{branch1}-\ref{branch3}), then bootstrap results from the two different implementations will show differences.
 
In summary, the positivity structure in $SU(N_f)$ adjoint crossing equations is not different from that of $SO(N_f^2-1)$ vector crossing equations. To bootstrap the non-$SO(N)$ symmetric theories like conformal QED$_3$, it is necessary to introduce gaps in the bootstrap equations which break the $SO(N_f^2-1)$ symmetry explicitly. Nevertheless, it is not clear whether the bootstrap bounds can converge to a non-$SO(N)$ symmetric theory just by introducing certain gap assumptions. In our bootstrap study of conformal QED$_3$, we introduce gap assumptions consistent with QED$_3$ spectrum and compare the bounds with $1/N_f$ perturbative results of 
conformal QED$_3$. Specifically, in the three sectors $\Ad_{\ell=0}, A\bar{A}_{\ell=0},T\bar{T}_{\ell=0}$, the lowest scalars are given by the four-fermion operators and their scaling dimensions are computed to subleading order (\ref{4fm}). Bootstrap solutions on the $SU(N_f)$ singlet bounds are $SO(N_f^2-1)$ symmetric and spectra in the three sectors are the same, which is true  at the leading order in conformal QED$_3$: $\Delta_{\Ad}=\Delta_{A\bar{A}}=\Delta_{T\bar{T}}=4$. The $SO(N_f^2-1)$ symmetry breaking factors appear at the subleading order. The question studied in Fig. \ref{SU20} is that after we introduce these $SO(N_f^2-1)$ symmetry breaking factors in the bootstrap equations, i.e., we require the linear functionals $\Vec{\beta}$ satisfy positive conditions
\be
\Vec{\beta}\cdot \cMad 
=\left( \alpha^+_{\mathbf{1}}, \alpha^+_{\Ad}, \alpha^-_{\Ad},
\alpha^-_{T\bar{A}},  \alpha^+_{A\bar{A}}, \alpha^+_{T\bar{T}}\right)\succeq 0_{1\times 6}, \hspace{1cm} \forall \Delta\geqslant \Delta_{R_i, \ell}^*\,. 
\ee
and for the gap assumptions $\Delta_{R_i, \ell}^*$, in most of sectors we adopt the unitary bounds, while for the three four-fermion scalars in $\Ad_{\ell=0}, A\bar{A}_{\ell=0},T\bar{T}_{\ell=0}$, we set
\bea
\Delta &\geqslant & \Delta_{\Ad,\ell=0}^*=\Delta_4+\frac{8(25- \sqrt{2317})}{3\pi^2 N_f},\\
\Delta &\geqslant &\Delta_{A\bar{A},\ell=0}^*=\Delta_4-\frac{64}{\pi^2N_f}, \\
\Delta &\geqslant&\Delta_{T\bar{T},\ell=0}^*=\Delta_4+\frac{64}{3\pi^2N_f},
\eea
and bootstrap the upper bound on the scaling dimension $\Delta_4$.
Another important $SO(N_f^2-1)$ symmetry breaking factor is from $T\bar{A}_{\ell=1}$ sector, in which the lowest spin 1 operator has scaling dimension $5\pm O(1/N_f)$. In Fig. \ref{SU20} and Appendix \ref{GapsD4}  effects of gaps in this sector have been studied. 

For the linear extrapolations in Table \ref{slrule}, we have computed $\Delta_4$ upper bounds to the numerical precision $10^{-5}$ with $\Delta_{\ad}$ fixed at the large $N_f$ perturbative result \cite{Gracey:1993sn}
\be 
\Delta_{\ad}=2-\frac{64}{3\pi^2 N_f}+\frac{256(28-3\pi^2)}{9\pi^4N_f^2}.
\ee 
The raw data is provided in an attached {\it Mathematica} file.

\section{Two adjacent kinks in the $SO(N)$ singlet bound?} \label{appendix:app2}
In this section we study fine structures of the new family of kinks in the bounds on $SO(N)$ singlet scalars.
As shown in the zoomed-in subplot in  Fig. \ref{SU20}, the new family of kinks in the $SO(N)$ singlet bounds may have fine structures with two nearby kinks for large $N$s. 
Such fine structure is not shown or indistinguishable in the singlet bounds with  small $N$s, see e.g. Fig. \ref{ONsig}. 
In  the $SO(20^2-1)$ singlet bound shown in Fig. \ref{SU20}, there is a prominent first kink near $\Delta_\phi \sim1.9$, followed by another mild kink near $\Delta_\phi \in(1.94,1.95)$. Comparing with the jump in the bound on the scaling dimension of the lowest $O(20^2-1)$ traceless symmetric scalar $\Delta_T$, we notice that the second kink in the singlet bound has scaling dimension $\Delta_\phi$ close to the top of the jump, while $\Delta_\phi$ of the first kink, which is close to $N_f=20$ QED$_3$ relates to the bottom of the jump.

\begin{figure}
\includegraphics[width=0.6\linewidth]{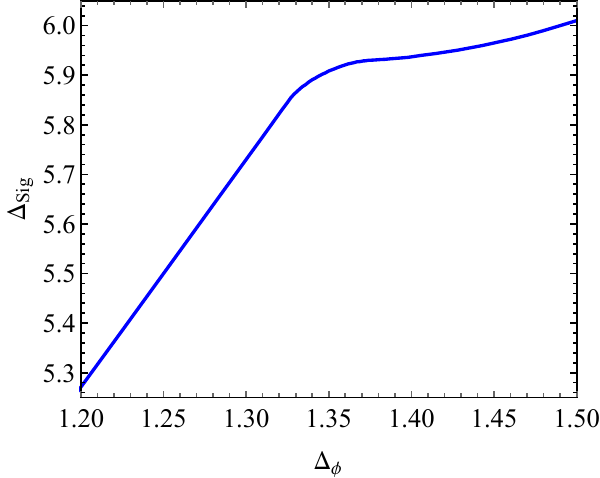}
\caption{Bound ($\Lambda=31$) on the scaling dimension of the lowest $SO(4^2-1)$ singlet scalar.} \label{SU4}
\end{figure}

In Fig. \ref{SU4} we provide another example for the kink in the $SO(15)$  singlet bound. The $SO(15)$ vector bootstrap bounds on the singlet and traceless symmetric scalars have been shown in Fig. \ref{lgNs} (purple lines). Fig. \ref{SU4} shows the zoomed in singlet bound near the kink. Comparing with kinks with small $N$s in Fig. \ref{ONsig}, the kink(s) in Fig. \ref{SU4} spread in a notable region with two transitions. In the bound on $\Delta_{T}$, $N=15$ is not large enough to form a sharp jump, while the range of $\Delta_\phi$ of the singlet kink(s) in Fig. \ref{SU4} is close to the kink in the bound on $\Delta_{T}$.


\section{ More discussions on the gap $\Delta_1^*$ in the spin $1$ $T\bar{A}$  sector} \label{gapinTAb}
\begin{figure}
\includegraphics[width=0.6\linewidth]{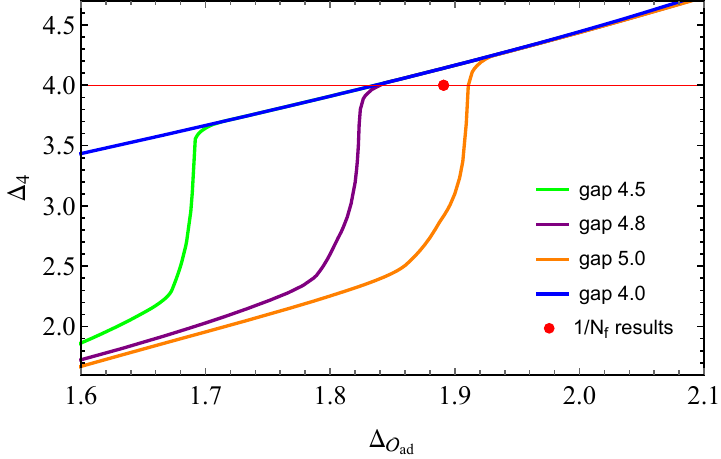}
\caption{Bounds on $\Delta_4$ with different gap assumptions $\Delta_1^*$ on the lowest spin $1$ operator in the $T\bar{A}$ sector.} \label{GapsD4}
\end{figure}
In Fig. \ref{SU20} we have shown that with  gaps $\Delta_1^*=4.5,4.8, \textrm{or } 5.0$ on the spin $1$ operators in $T\bar{A}$ sector, bootstrap bounds on $\Delta_4$ can form sharp jumps whose positions depend on the specific values of $\Delta_1^*$. In this section we provide more discussions on the bootstrap results. We reproduce bootstrap bounds on $\Delta_4$ in Fig. \ref{GapsD4} for convenience.

The lowest spin 1 operator in $T\bar{A}$ sector has scaling dimension $5\pm 1/N_f$. The subleading order correction is not known yet. The physical gap $\Delta_1^*$ could be slightly above or below $5$ for $N_f=20$, depending on the sign of the subleading order correction. An interesting observation in Fig. \ref{GapsD4} is that with a gap $\Delta_1^*=5$, the large $N_f$ predictions on QED$_3$ $(\Delta_{\ad}, \Delta_4)\simeq(1.891,4.0)$ are excluded! Note $\Delta_{\ad}\simeq 1.891$ is given by $1/N_f$ perturbative result at the order $1/N_f^2$, which is expected to be well close to the physical spectrum. $\Delta_4\simeq 4$ is given by  large $N_f$ expansions on the scaling dimensions of the four-fermion operators \ref{4fm}, which can be shifted by higher order corrections while it is unlikely to be lowered into the bootstrap allowed region in Fig. \ref{GapsD4}: $\Delta_4<3$ at $\Delta_{\ad}=1.891$. We expect the $1/N_f$ predictions on QED$_3$ are excluded due to the gap assumption $\Delta_1^*=5$. The physical gap $\Delta_1^*$ should be smaller than 5 and the subleading order correction is negative.

Moreover, with a gap $\Delta_1^*=4.8$ the top of the jump is close to the physical value $\Delta_4=4$. Near the jump we have $\Delta_{\ad}\simeq 1.84$, smaller than the large $N_f$ prediction $\Delta_{\ad}\simeq 1.89$. As shown in the linear extrapolation of bound on $\Delta_4$, the large $\Lambda$ extrapolation will help to reduce the discrepancy. 
It is interesting to compare with subleading order corrections on the four-fermion scalars (\ref{4fm}). For $N_f=20$ QED$_3$ the formulas in  (\ref{4fm}) give anomalous dimensions $\Delta_R^{(1)}$:
\be
(\Delta_{\Ad}^{(1)}, \Delta_{A\bar{A}}^{(1)}, \Delta_{T\bar{T}}^{(1)} )\simeq
(-0.31, -0.32, 0.11), \label{4fm-1}
\ee 
close to the ``anomalous dimension" $-0.2$ given by the gap  $\Delta_1^*=4.8$. A critical question is when the gap $\Delta_1^*$ is close to the physical spectrum of the lowest spin 1 operator in $T\bar{A}$, will the top of the jump in the bound on $\Delta_4$ get close to QED$_3$? It is also interesting to isolate QED$_3$ solutions near the jumps using single correlator bootstrap \cite{Li:2017kck}. We will provide a more comprehensive study for this problem in the near future.

It is quite encouraging that with a suitable gap in $T\bar{A}$ sector, bootstrap bound can form a sharp jump close to the physical spectrum. On the other hand, our results disclose a subtle challenge to solve conformal QED$_3$ using conformal bootstrap. The bootstrap bounds, in particular the kinks depend on gaps imposed in the bootstrap equations. Therefore one has to resort to the results from large $N_f$ expansions or other approaches to obtain bootstrap bounds relevant to QED$_3$. Reliable information on QED$_3$ is not available for physically interested theories with small $N_f$ and our hope is to solve QED$_3$ without specific information of the theory. It is important to find new ingredients to resolve the gap-dependence problem for future bootstrap studies.

\section{Lower bounds on $c_J$ and $c_T$ in $SU(20)$ adjoint bootstrap} \label{appendix:app3}

\begin{figure}
\includegraphics[width=0.45\linewidth]{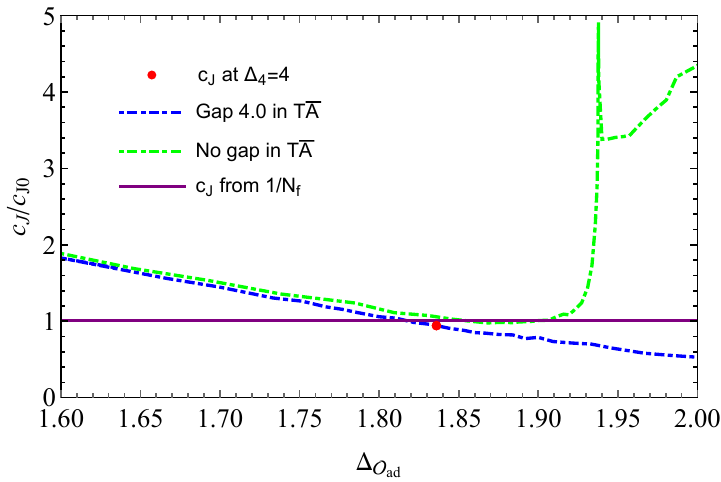}
\includegraphics[width=0.46\linewidth]{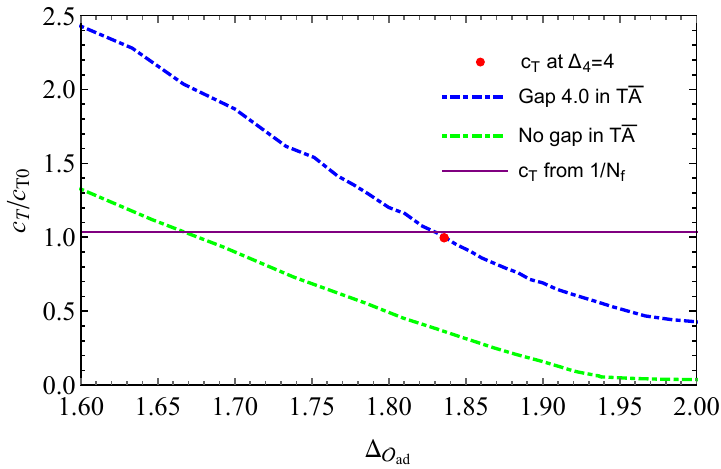}
\caption{Lower bounds on central charges $c_J$ (left panel) and $c_T$ (right panel) near the upper bounds on $\Delta_4$ with/without a gap $\Delta_1^*=4$ in Fig. \ref{SU20}. The central charges are given with the normalization in which $c_J=c_T=1$ for $N_f=20$ two-component free fermions. The sharp pike in the $c_J$ lower bound without $T\bar{A}$ gap (green line in the left panel) can be affected by our sample points near the jump in $\Delta_4$ bound in Fig. \ref{SU20}. Clearly there is a jump in $c_J$ bound near $\Delta_{\ad}=1.94$, but the shape of the bound right to the jump may change notably if our sample points get more close to the boundary.} \label{cJcT}
\end{figure}

In Fig. \ref{SU20} we have shown that the upper bound (blue line) on $\Delta_4$ with gap assumptions breaking $SO(20^2-1)$ symmetry converges to the $1/N_f$ perturbative results of $N_f=20$ QED$_3$ in the large $\Lambda$ limit, which suggests that conformal QED$_3$ could provide a nearly extremal solution to the bootstrap bound with non-$SO(20^2-1)$ symmetric gap assumptions, up to uncertainties from linear extrapolation.
In this section, we provide more evidence for the relation between the $SU(20)$ adjoint bootstrap bounds and conformal QED$_3$. We will study bootstrap bounds on the $SU(20)$ conserved current central charge $c_J$ and stress tensor central charge $c_T$ with $\Delta_4$ fixed near its upper bounds. 


Central charges $c_J$ and $c_T$ play important roles in the dynamics of local interacting CFTs. For QED$_3$, these two parameters have been computed using $1/N_f$ perturbative method to subleading order \cite{Giombi:2015haa}
\bea \label{cclargeNf}
c_J &=c_{J0} \left(1+\frac{0.1429}{N_f}+O(1/N_f^2)\right), \\
c_T &=c_{T0} \left(1+\frac{0.7193}{N_f}+O(1/N_f^2) \right),  
\eea
where $c_{J0}$ and $c_{T0}$ are the central charges from $N_f$ flavors of two-component free fermions. The central charges $c_J$ and $c_T$ are especially helpful to distinguish Abelian gauge theories like QED$_3$ from Yang-Mills gauge theories. The low lying gauge invariant operators constructed from matter fields are similar in these theories and it may be hard to distinguish QED$_3$ from Yang-Mill gauge theories with the same flavor symmetry, e.g. 3D Quantum chromodynamics with $SU(N_c)$ gauge symmetry (QCD$_3$). For instance, the $SU(N_f)$ adjoint fermion bilinears $\ad$ appear both in QED$_3$ and QCD$_3$, and their scaling dimensions are the same at leading order with possibly different higher order corrections.
In contract, central charges $c_J$ and $c_T$ are proportional to dimensions of the gauge group representations constructed by fermions,
which are significantly different in QED$_3$ and QCD$_3$. Therefore bounds on $c_J$ and $c_T$ can persuasively show whether the underlying theory relates to QED$_3$ or QCD$_3$.

In Fig. \ref{cJcT} we show lower bounds on $c_J$ and $c_T$ with $\Delta_4$ fixed near its upper bounds (blue and green lines  in Fig. \ref{SU20}). On the bound of $c_J$ without gap in $T\bar{A}$ sector, there is a sharp jump near $\Delta_{\ad}=1.94$, corresponding to the jump in the bound on $\Delta_4$, given by the green line in Fig. \ref{SU20}. In this work, we will be particularly interested in the bounds near the physical spectrum of QED$_3$ (\ref{4fm}): $\Delta_4\simeq4$, which intercepts the $\Delta_4$ upper bound (blue line) in Fig. \ref{SU20} at $\Delta_{\ad}\sim 1.836$.  
The scaling dimension $\Delta_{\ad}\sim 1.836$ with $\Delta_4=4$ is obtained at $\Lambda=31$, which is lower than the $1/N_f$ perturbative result $\Delta_{\ad}\sim 1.891$. Using  linear extrapolations, the two estimates  become reasonably close with each other in the large $\Lambda$ limit, as shown in Fig. \ref{SU20}
and Table. \ref{slrule}. Remarkably, near $\Delta_4=4$, lower bounds on $c_J$ and $c_T$ given by the red dots in Fig. \ref{cJcT} are quite close to the large $N_f$ perturbative results (\ref{cclargeNf}). Note that both $c_J$ and $c_T$ change drastically with $\Delta_{\ad}$, it is highly non-trivial that bounds on $c_J$ and $c_T$ get close to their physical value just near $\Delta_4=4$. As we have discussed before, $c_J$ and $c_T$ in Yang-Mills gauge theories, like QCD$_3$ have $c_J$ and $c_T$ about $N_c$ times larger than (\ref{cJcT}). Therefore the underlying theory near the bounds on $c_J$ and $c_T$ at $\Delta_4=4$ should be QED$_3$ or analogous 3D Abelian gauge theories instead of Yang-Mills gauge theories.
\end{document}